\documentclass[pra,twocolumn,showpacs,preprintnumbers,amsmath,amssymb,superscriptaddress, eqsecnum]{revtex4}
\usepackage{bm,graphicx,amsbsy,amsmath,amsfonts,amsthm,color,mathrsfs}

\begin{document}

\title{Low-loss nonlinear polaritonics}

\author{Sergey A. Moiseev}
\address{Kazan Physical-Technical Institute, Russian Academy of Sciences, Kazan 420029, Russia}
\address{Institute for Quantum Information Science, University of Calgary, Alberta T2N 1N4, Canada}
\author{Ali A. Kamli}
\address{Department of Physics, King Khalid University, Abha, 61314 Saudi Arabia}
\address{National Centre for Mathematics and Physics, KACST, Riyadh 11442, Saudi Arabia}\address{Institute for Quantum Information Science, University of Calgary, Alberta T2N 1N4, Canada}
\author{Barry C. Sanders}
\address{Institute for Quantum Information Science, University of Calgary, Alberta T2N 1N4, Canada}

\date{\today}

\begin{abstract}
We propose large low-loss cross-phase modulation between two coupled surface polaritons
propagating through a double electromagnetically-induced transparency medium
situated close to a negative-index metamaterial.
In particular a mutual $\pi$ phase shift is attainable between the two pulses at the single photon level.
\end{abstract}

\maketitle

\section{Introduction}
\label{sec:introduction}
A low-loss nanoscale all-optical switch~\cite{VGX08} could revolutionize photonics through its compatibility with
proposed nanophotonic structures, speed, and efficacy at low light levels.
Although such a device is needed, its creation has been prevented by the poor trade-off between confinement of
light and losses and the incompatibility of low light levels with strong Kerr nonlinearity.
Some of these challenges may be ameliorated by ongoing research.
For example giant cross-phase modulation (XPM) could be enabled by double electromagnetically induced transparency (double EIT or `DEIT')~\cite{LI01,WMS06, MCL08}.
Furthermore surface plasmons~\cite{Mai07} could exploit subwavelength optics~\cite{BDE03},
albeit with large losses that may be lessened by clever strategies~\cite{NTF04,GP06,AMY+07}.

We show that, by coupling two surface polaritons (SPs) in a DEIT medium situated close to
an interface between a dielectric and a negative-index metamaterial (NIMM)~\cite{EZ06,Sha07},
such that material properties are judiciously chosen~\cite{KMS08}, giant cross-phase modulation between the two SPs can be achieved
in a low-loss, sub-wavelength confinement regime.
In particular a mutual $\pi$ phase shift between the two pulses is attainable for weak fields
with a mean photon number of one, thereby opening the prospect of
deterministic single-photon quantum logic gates for quantum computing~\cite{KMN+07}.

The mutual phase shift between two pulses is achieved by creating a Kerr-nonlinear
refractive index simply expressed as $n=n_0+n_2I$ for $n_0$ the linear refractive
index, $n_2$ the coefficient for the nonlinear index, and~$I$ the intensity of the light field.
For two separate field modes~$a$ and~$b$,
the mutual phase shift of~$a$ due to~$b$ is phase difference experienced by~$a$ due to
its interaction with~$b$ compared to the phase shift if~$b$ were turned off.
Similary, the mutual phase shift of~$b$ is the phase difference experienced by that mode
for~$a$ on vs the case of~$a$ off.

In natural media, $n_2$ is quite small.
Furthermore the mutual phase shift is proportional not only to $n_2$ but also
to the energy density of the field and the interaction time between the two pulses.
Typically interaction time is quite short,
and energy density is diffraction-limited.
Fortunately, double electromagnetically-induced transparency combined with strongly
driven cross-phase modulation simultaneously creates a large $n_2$ nonlinearity,
compresses the energy of the pulse in the direction of propagation,
and increases the interaction time by slowing or even stopping the pulses~\cite{LI01,WMS06}.

However, the resultant mutual phase shift is expected to be at best on the order of
$10^{-5}$ radians per photon squared.
In other words, for modes $a$ and~$b$ with mean photon number of one,
the mutual phase shift is only around $10^{-5}$ rad, which is far too small
for weak-field all-optical phase-triggered switches.

One way to boost the nonlinear phase shift is to compress the field energy in the
transverse direction.
Then all the ingredients are in place for huge mutual phase shifts between pulses~$a$ and~$b$.
Transverse confinement is made possible by bringing a second medium in close proximity to the
interaction region and driving this second medium.
In this case, exponential confinement of the second medium's evanescent field can produce the desired strong confinement, but there is a problem:
some of the electromagnetic field energy penetrates into the second medium,
whereas almost all the field energy should be in the first medium for weak-field phase-triggered all-optical switches to work.

Field penetration in medium~2 can be restricted for this medium being a metal:
the field is then converted to a plasmon with an evanescent field that acts on the nonlinear
medium located above medium~2.
Using a metal has the serious drawback, though,
that the plasmon is notoriously lossy.
This loss must be avoided for switching to be efficient.

Recently we discovered that making medium~2 a negative-index metamaterial (NIMM)
rather than a dielectric or a metal combines the best field-confinement features of both.
Specifically it is possible to minimize losses for selected  exponential confinement of the field~\cite{KMS08}.
Thus an electromagnetic pulse excites a ``low-loss surface polariton''~(LLSP)
in this NIMM, for which
dielectric permittivity~$\varepsilon_0\varepsilon(\omega)$
and magnetic permeability $\mu_0\mu(\omega)$ are both negative
with~$\omega$ the LLSP mode carrier frequency and
both $\varepsilon$ and $\mu$ dimensionless.

Here we show that two LLSPs will interact via the Kerr nonlinear medium,
retain their low-loss nature, and yield large mutual phase shifts (e.g., $\pi$ radians).
Thus, two electromagnetic pulses can effect strong mutual phase shifts,
as shown in Fig.~\ref{fig:NPscheme},
by converting to LLSPs, then interacting via a nonlinear medium in medium~1, followed by converting
back to electromagnetic pulses.
Each of the two media has permittivity~$\varepsilon_j$ and permeability~$\mu_j$,
with~$j=1$ for the upper ($z>0$) dielectric medium and $j=2$ for the lower ($z<0$) NIMM medium.
We refer to our proposal for strong nonlinear interactions between LLSPs as \emph{`low-loss nonlinear polaritonics'}.

\section{Atoms and fields near the interface}
\label{sec:atoms}
\begin{figure}
 \centering
 \includegraphics[width=80mm]{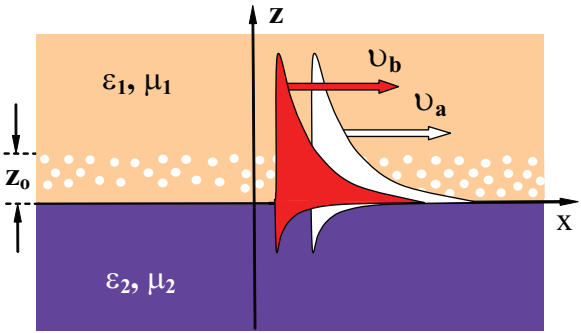}
 \caption{
 	(color online)
	Two SPs, shown as dark (red) and white pulses created between two media $i=1,2$
	with permittivities~$\varepsilon_i$ and permeabilities~$\mu_i$.
	Medium~1 is in the region~$z>0$ and medium~2 in the region $z<0$.
	The pulses propagate forward in the $+x$ direction and are exponentially
	confined to the interface~$z=0$.
	The white spots represent multi-level atoms in the double $\Lambda$ configuration and
	are confined to a region between $z=0$ and $z=z_0$.
	}
\label{fig:NPscheme}
\end{figure}

A collection of multi-level atoms is located in medium~1 in close proximity with the interface between the two media,
as depicted in Fig.~\ref{fig:NPscheme}.
Only five levels are required for our proposed DEIT scheme so we refer to these multi-level atoms as five-level atoms (5LAs).
These atoms could form a cold gas (e.g. $^{87}$Rb) or a solid-state medium (e.g. Pr:YSiO).
DEIT has been demonstrated in the former~\cite{MCL08}
 and EIT in both gas and solid systems~\cite{HH99,Tur01}.
The 5LA structure allows the creation of LLSPs, realizes DEIT for slowing the beams,
and enables huge cross-phase modulation as we shall now see.

In our analysis, we treat the field modes as propagating nearly plane waves, which
provide a convenient basis for describing pulses,
and there is translational invariance
in both the $x$- and $y$-directions, except, of course, at the planar dielectric-NIMM interface.
The LLSP is confined in the $z$ direction because of evanescence,
and confinement in the $x$ direction is due to DEIT.


To analyze low-loss nonlinear polaritonics,
we  require the recently obtained LLSP dispersion relation~\cite{KMS08}.
The dielectric (medium~1) is assumed to have constant homogeneous~$\varepsilon_1$ and~$\mu_1$,
and, for the NIMM,
\begin{align}
	\varepsilon_2(\omega)
		\equiv \varepsilon_{\rm r}(\omega)+{\rm i}\varepsilon_{\rm i}(\omega)
		= \varepsilon_b-\frac{\omega_{\rm e}^2}{\omega(\omega+{\rm i}\gamma_{{\rm e}})}, \\
	\mu_2(\omega)
		\equiv \mu_r(\omega)+{\rm i}\mu_m(\omega)
		=\mu_b-\frac{\omega_{\rm m}^2}{\omega(\omega+{\rm i}\gamma_{{\rm m}})},
\end{align}
where $\omega_{\rm e}$ ($\gamma_{\rm e}$) corresponds to the electric plasma frequency (decay rate)
and $\omega_{\rm m}$ ($\gamma_{\rm m}$)
to the magnetic plasma frequency (decay rate)~\cite{EZ06,Mai07}.
Typical values are $\omega_{\rm e}=1.37 \times 10^{16}\text{s}^{-1}$ and $\gamma_{\rm e}=2.73 \times 10^{13}\text{s}^{-1}$, and we assume for the magnetic components
$\omega_{\rm m}=\omega_{\rm e}/6$ and $\gamma_{\rm m}=\gamma_{\rm e}/1000$.
The background dielectric constant $\varepsilon_b$ in real metals~\cite{Mai07} is between~$1$ and~$10$.
In our analysis we fix $\varepsilon_b=2$ and $\mu_b=2$.

The complex wave number along the $+x$ axis for a plane-wave SP mode along the
$x-y$ plane is denoted
\begin{equation}
\label{eq:K||}
	K_{\parallel}=k_{\parallel}(\omega)+\text{i}\kappa(\omega)
\end{equation}
for $k_{\parallel}$ and $\kappa$ the real and imaginary parts, respectively.
The normal component~$k_j$ of the SP wavevector in each region is related to $K_\parallel$ by
\begin{equation}
\label{eq:kj2}
	k_j^2=K_\parallel^2-\omega^2\varepsilon_j\mu_j/c^2,\;
	j=1,2.
\end{equation}
The wavenumbers on each side of the interface ($z=0$)
are related by the boundary conditions~\cite{Mai07}
so
\begin{equation}
	-k_2/k_1= \eta\equiv\eta_\varepsilon := \varepsilon_2/\varepsilon_1
\end{equation}
for a transverse magnetic (TM) SP, whereas, for the transverse electric~(TE) SP,
\begin{equation}
	\eta\equiv\eta_{\mu}:=\mu_2/\mu_1.
\end{equation}

Using relations~(\ref{eq:kj2}),
$k_1$ and $k_2$ can be eliminated.
We then obtain the complex wavevector
\begin{equation}
\label{eq:TM}
	K_{\parallel}=\frac{\omega}{c}
		\sqrt{\varepsilon_2\mu_2\frac{1-\eta_{\varepsilon}/\eta_{\mu}} {1-\eta_{\varepsilon}^2}},
\end{equation}
for the TM mode with
the real part giving dispersion and the imaginary part giving absorption loss.
The absorption tends to $\kappa(\omega) \rightarrow0$
if $\gamma_{\rm e} \rightarrow0$ and $\gamma_{\rm m}\rightarrow 0$.
The TE mode case is obtained by exchanging $\eta_{\varepsilon}\leftrightarrow\eta_{\mu}$.

The dispersion relation and boundary condition yield SP amplitude vs distance~$z$
from the interface.
For the NIMM-dielectric system, the field is exponentially confined in the $z$-direction with amplitude given
(for the TM polarized case) by
\begin{equation}
\label{eq:field}
	\left|E_{0,j}({\bm r,k_\parallel)}\right|
		\sim\frac{{\rm e}^{-\left|z\right|/\zeta_j}}{\sqrt{L_{z}}} \exp\left[-\kappa(\omega) x\right], j=1,2.
\end{equation}
with characteristic mode length
\begin{align}
\label{eq:Lz}
	L_{z}=&\left[\tilde \varepsilon_1 \left(1+\frac{|k_\parallel|^2}{|k_1|^2}\right)+\frac{\omega^2}{c^2}
        \tilde \mu_1 \frac{|\varepsilon_1|^2}{|k_1|^2} \right] \zeta_1
        			\nonumber	\\	&
		+\left[\tilde \varepsilon_2 \left(1+\frac{|k_\parallel|^2}{|k_2|^2}\right)+\frac{\omega^2}{c^2}
        \tilde \mu_2 \frac{|\varepsilon_2|^2}{|k_2|^2} \right] \zeta_2,
\end{align}
where
\begin{equation}
	\tilde f_j:=\text{Re}\left[\frac{\partial(\omega f_j)}{\partial \omega}\right]
\end{equation}
and confinement
\begin{equation}
 	\zeta_j \approx \frac{1}{\text{Re}( k_j(\omega))},\, j=1,2.
\end{equation}
In the dielectric $+z$ region
\begin{equation}
	\zeta_1\approx \frac{c}{\omega\text{Re}\sqrt{\varepsilon_1\mu_1
		\frac{1-\eta_{\varepsilon} \eta_{\mu}}
{\eta_{\varepsilon}^2-1}}},
\end{equation}
which characterizes the scale for confinement in the dielectric.
We shall see in the next section that these three quantities,
namely losses, confinement and mode length,
can be optimized to maximize the field amplitude which is important for the nonlinear polaritonics.

\section{Cross-phase modulation}
\label{sec:cross-phase}

\begin{figure}
 \centering
 \includegraphics[width=80mm]{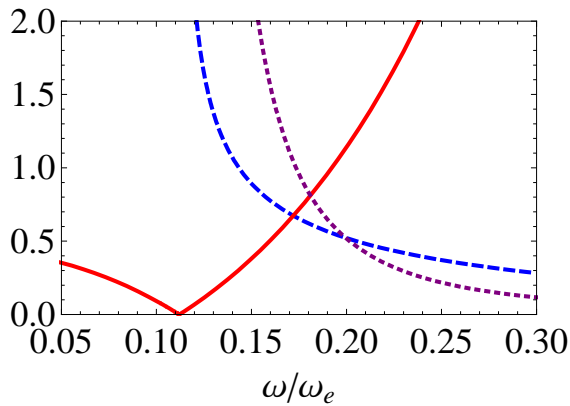}
 \caption{
 	(color online)
 	Characterizing LLSPs as a function of $\omega/\omega_{\rm e}$,
	with~$\omega_{\rm e}=1.37 \times10^{16} \text{s}^{-1}$, for
	confinement factor~$\zeta_1$ (blue-dashed), and
	mode length (purple-dotted) normalized to the Rb D$_2$ wavelength $\lambda=780{\rm nm}$
	and absorption (red-solid) ($\times 5.10^3$m$^{-1}$).}
\label{fig:losses}
\end{figure}

Fig.~\ref{fig:losses} demonstrates spectral dependence of the absorption coefficient~$\kappa$ and spatial confinement~$\zeta_1$ of SP modes for the NIMM-dielectric interface in the spectral range close to $\omega_0$ of complete suppression of the losses.
As seen from Fig.~\ref{fig:losses},  a complete suppression of losses  is accompanied by a deconfinement
of LLSP modes (i.e.\ $\kappa\rightarrow0$ with~$\zeta\rightarrow\infty$).
One way to understand this trade-off between confinement and losses
is from energy considerations:
strengthening confinement of the surface polariton on the dielectric side increases
the fraction of electromagnetic energy on the NIMM side of the interface.

\begin{figure}
 \centering
\includegraphics[width=80mm]{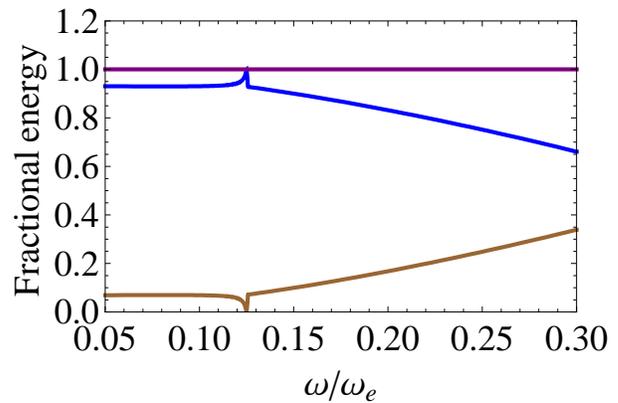}
 \caption{
 	(color online)
 	The fractional energies in the NIMM part (lower brown curve), in the dielectric part
	(middle blue line) and total energy (upper purple line)
        as functions of $\omega/\omega_{\rm e}$.}
\label{fig:new}
\end{figure}

Energy transport at optical frequencies on the NIMM side involves scattering of free electrons hence
large losses. Fig.~\ref{fig:new} depicts the fractional energies in the dielectric and NIMM parts.
For complete suppression of loss,
which occurs for a frequency designated by~$\omega_0$,
the field energy resides completely in the dielectric but with poor confinement.

Frequency~$\omega_0$ and  decay constants~$\gamma_{\rm e}$ and $\gamma_{\rm m}$
can be chosen by judiciously selecting the properties of
the NIMM and controlling external parameters such as temperature.
Fig.~\ref{fig:losses} shows that the ratio $L_z/\lambda\sim 60$ yields a minimum for spatial extent of the
field in the dielectric medium, but a smaller geometric transverse spatial size $L_z/\lambda<10$ produces similarly confined SP modes up to a factor of $\sim10^{-3}$.

\begin{figure}
 \centering
\includegraphics[width=40mm]{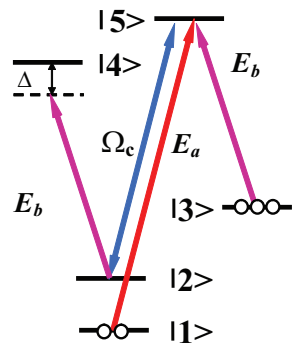}
\caption{
	(color online)
	Frequencies of two slow LLSP fields~$E_{a,b}$, of the control LLSP field~$\Omega_c$
	and energy diagram for the D$_2$ line of $^{87}$Rb.}
\label{fig:6LA}
\end{figure}

Now we see how these confined LLSPs will interact in a Kerr nonlinear
medium embedded in the dielectric.
In particular we focus on the $^{87}$Rb gas system since its D$_2$ transition wavelength of 780nm
appears to be commensurate with current NIMM technology~\cite{DWSL07}.
Moreover the $^{87}$Rb D$_2$ line yields DEIT~\cite{MCL08} and is expected to yield XPM~\cite{WMS06}.
These phenomena are dominated by five of the D$_2$ lines:
the energy diagram and level scheme are shown in Fig.~\ref{fig:6LA}.

\begin{figure}
\includegraphics[width=80mm]{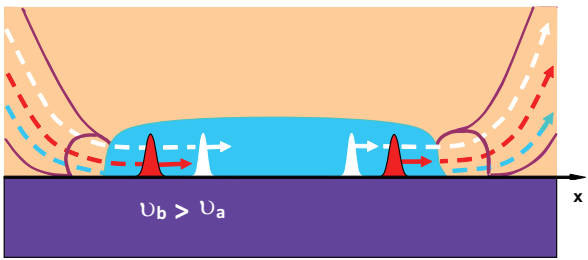}
\caption{
 	(color online)
	Spatial and temporal diagram of excitation and interaction of two slow LLSP pulses propagating
	with group velocities~$v_b>v_a$,
	with the $a$ pulse in white and the $b$ pulse in dark red,
	in the presence of a control field~$\Omega_c$ shown in light blue.
	The second (dark red) LLSP pulse outraces the first (white) LLSP pulse in the medium output.
	The dashed lines on the left and right indicate conceptually how the fields are directed
	into and out of the interaction region, and the solid lines on other side represent the
	waveguide that brings the field in and out. 
	}
\label{fig:two_pulse}
\end{figure}

Let us assume that the two interacting LLSP pulses are excited one by one at the interface input with slightly different adjusted group velocities~$v_{a,b}$. Let the second LLSP pulse have larger group velocity $v_b>v_a$ and outrace the first LLSP at the medium output as depicted in Fig.~\ref{fig:two_pulse}.

We derive  nonlinear coupled equations for two slowly propagating LLSP fields by taking into account spatial confinement of interaction with resonant atomic systems. We note that  LLSP  $E_b(t,x)-$
modes experience different strengths of the nonlinear interaction compared with other $a$ and $b$ modes
in the transverse $y \times z$ cross-section due to highly inhomogeneous intensities of the electromagnetic fields.

All these nonlinear interactions within the cross-section do not alter the usual form of the nonlinear equation for the traveling probe pulse with amplitude $E_b$:
\begin{align}
\label{eq:nonlin}
	\Bigg(\frac{1}{v_b}\frac{\partial}{\partial t} +& \frac{\partial}{\partial x}
			- i\frac{\partial^2}{2 k_{\parallel}\partial y^2}\Bigg) E_b
				\nonumber	\\
		&= i\left[\chi_a^{(3)} I_a+\chi_b^{(3)} I_b \right] E_b,
\end{align}
leading only to averaged nonlinear Kerr coefficients $\chi_a^{(3)}$ and $\chi_b^{(3)}$, where
\begin{equation}
	I_{a,b} = |E_{a,b}(t,x,y)|^2.
\end{equation}
The term proportional to~$\chi_b^{(3)}$ describes self-phase modulation (SPM) of the~$E_b(t,x,y)$ field.

SPM accompanies the XPM effect with a similar magnitude,
and SPM can lead to unwanted effects:
temporal distortion of the pulse including chirping~\cite{ZS72,AE81},
squeezing the amplitude quadrature~\cite{BH91},
and quantum limits to all-optical switching with coherent states in Kerr media~\cite{SM92a,SM92b,MdM92}.
For XPM the problem of temporal distortion and chirping can be significantly alleviated by
beam shaping to enable a uniform phase across one of the two pulses in the medium~\cite{MWMS10},
and switching limits can be compensated interferometrically~\cite{SM92b,MdM92}.
If SPM is compensated then $\chi_b^{(3)}$ in Eq.~(\ref{eq:nonlin}) can be effectively neglected.

In the transverse $y$-direction, the beams are focused by lenses so confinement is diffraction-limited in this dimension.
The problem of beam spreading is not significant because the propagation length for LLSPs is small.
The diffraction-limit in the $y$ direction could be reduced by inserting defects into
the multi-level atomic medium\cite{Mai07} or by creating a surface groove in the interface~\cite{BVDE05},
so we can ignore term proportional to
\begin{equation}
	\frac{1}{2 k_{\parallel}}\frac{\partial^2}{\partial y^2}
\end{equation}
in Eq.~(\ref{eq:nonlin}) as well.

When the two pulses pass through a 5LA system  of Rb gas of spatial thickness $z_0$ in the $z$ direction,
the resultant Kerr nonlinearity is
\begin{align}
\label{eq:nonlinear}
	\chi _{\rm a}^{(3)}=&\frac{2\pi n_{1} z_0}{\hbar^4 v_{b,0} \vert \Omega _c \vert ^2\Delta}
		\Phi\left[\left(\tilde k_a^{\rm p} + \tilde k_b^{\rm p} - \tilde k_{\rm c}\right)z_0\right]
			\nonumber	\\	&
		\times\left\langle {\left\vert \bm{d}_{24}\cdot\bm{E}_b\right\vert ^2\left\vert \bm{d}_{15}\cdot
		\bm{E}_a\right\vert ^2} \right\rangle
\end{align}
for
\begin{equation}
            \Phi(u)=\text{e}^{ -u}\frac{\text{sinh}u}{u}.
\end{equation}
Here the group velocity of the $l^{\rm th}$ ($l=a,b$) slowly propagating LLSP pulse in the presence of resonant atoms
is
\begin{align}
\label{groupvel}
      v_l = \frac{v_{l,0}}{1 + \beta _l},
\end{align}
where $v_{l,0} $ is the group velocity of the
$l^{\rm th}$ LLSP pulse in the absence of atoms.
Also we have
\begin{equation}
	\beta _b
		=2\pi n_3 z_0\Phi[(\tilde k_b^p - \tilde k^c z_0]
			\frac{\left\langle {\vert \bm{d}_{35}
			\cdot \bm{E}_b \vert ^2}\right\rangle }{\hbar ^2\vert \Omega _c \vert ^2}.
\end{equation}

The atomic transition dipoles are
$\bm{d}_{24}$ and $\bm{d}_{15}$ between levels~2 and~4 and
between~1 and~5, respectively, as shown in Fig.~\ref{fig:6LA}.
The classical electric field vector is $\bm{E}_{a,b}$ for the~$a$ and~$b$ LLSP pulses,
with energies corresponding to a single photon in each field, respectively.
Averaging over the orientation of the atomic dipole moments is included in the
expectation value $\left\langle {\cdots}\right\rangle$;
$z_{\rm 0}$ is spatial thickness of the atomic medium along $z$-direction, $\tilde k_{a,b}^{p}$ is the real part of the wave vector of the
SP field ($a$, or~$b$) along z and $\tilde k^{c}$ is the control field wave vector along z,
$\Omega _c $ is the Rabi frequency of the
control field, $n_{m}$ is atomic density on the $m$-th level, $\Delta $ is the spectral detuning.

Thus we see that, in comparison with free light fields, the nonlinear LLSP interactions demonstrate a robustness of the homogeneous phase shift in the cross-section similar to the propagation of light in the single mode waveguide.
The corresponding nonlinear phase shift experienced by the weak LLSP field~$b$,
after passing through the other weak LLSP pulse $a$,
is
\begin{equation}
	\varphi_b
		=\frac{1}{(1/v_a-1/v_b)} \frac{\chi_a^{(3)}}{v_{a,0}}
\end{equation}
The resultant phase shift on field~$b$ in a medium of length $L_x$, with
\begin{equation}
	L_x(1/v_a-1/v_b)\cong 2\tau
\end{equation}
for~$\tau$ the temporal duration of the LLSP pulse, is
\begin{equation}
	\varphi_b
		\cong\frac{L_{x}}{2\tau}\frac{\chi_a^{(3)}}{v_{a,0}}.
\end{equation}

For the 5LA $^{87}$Rb gas we assume ideal EIT conditions,
which take place for small enough thickness of the atomic layer~$\tilde k_j^{\rm p} z_0 \approx \tilde k^{\rm c} z_0 \approx1$.
The $^{87}$Rb $D_2$ line has a transition wavelength 780 nm~\cite{Ste} and the dipole moments for such transition of the order~$4{\rm e}a_0$ where $e$ is the electronic charge and $a_0$ is the Bohr radius.

We have chosen the media parameters such that the transition wavelength 780 nm corresponds
to the SP frequency $\omega=0.144\omega_{\rm e}$.
This frequency is quite close to $\omega_0$ where SP fields exhibit low losses and large confinement.
The linewidth is on the order of MHz, and the detuning is $\Delta=1.38$ MHz.
The Rabi frequency for the control field is $\Omega_c =1 $ MHz.
The atomic density of level~$1$ is taken to be $10^{14}\text{cm}^{-3}$ (typical gas),
and the medium size along the $x$-direction is $L_x\approx 0.3{\rm mm}$.

If the SP pulse temporal duration is of the order~$\tau\approx 1\mu s$,
the mean thermal velocity of $^{87}$Rb-atoms should satisfy the condition
\begin{equation}
\label{eq:upsilonRb}
	\upsilon_{\rm Rb}< 0.1\lambda \Delta/2 \pi
\end{equation}
for the Doppler broadened resonant line,
which limits the temperature of $^{87}$Rb gas to
\begin{equation}
	T< m_{\rm Rb} (\upsilon_{\rm Rb})^2/2k_{B}\approx 0.8 \mu K.
\end{equation}
It is possible to increase the minimal temperature by using solid-state media interfaces~\cite{WPZ06} with spectral tailoring of narrow resonant lines~\cite{Tur01}.
Here, the interface of diamond (containing resonant NV centers) with NIMM  looks quite promising based on recent experiments with fabricated diamond/metal interfaces~\cite{vdSHD+09}.
\begin{figure}
 \centering
 \includegraphics[width=80mm]{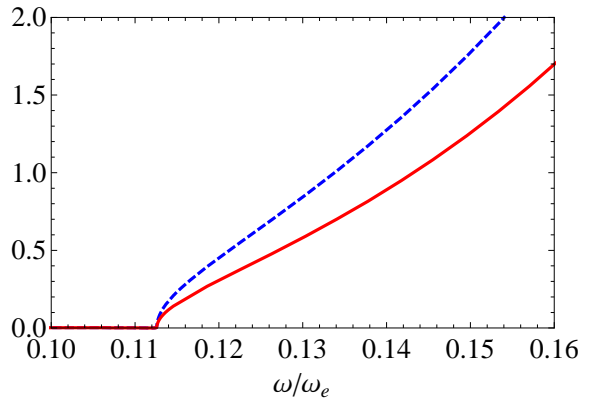}
 \caption{
 	(color online)
 	Third order susceptibility $\chi_{\rm a}^{(3)}$ (blue dashed line) due to
	double EIT scheme ($\times 10^5$) as a function of probe field frequency $\omega/\omega_{\rm e}$, and the corresponding  phase shift $\varphi_b$ (red solid line) due to SP cross-phase modulation in double EIT scheme (in units~$\pi$) as a function of probe field frequency $\omega/\omega_{\rm e}$.}
\label{fig:Phase_shift}
\end{figure}
For this set of parameters we show in  Fig.~\ref{fig:Phase_shift} the Kerr nonlinear coefficient for field~$b$, and the corresponding phase shift due to cross-phase modulation between the SP pulses.

\section{Large cross-phase modulation}
\label{sec:large}
Near and below the frequency $\omega_0$ where SP fields exhibit low losses, the real part of $k_1$ is nearly zero.
This leads to field deconfinement, namely poor confinement (large $\zeta_j$) and large mode length, which leads naturally to weak coupling of SP fields to atoms near the interface.
Thus, both the Kerr nonlinearity coefficient and the phase shift are small.
As we surpass this frequency region to higher frequencies where real part $k_1$ gets larger,
field confinement is further enhanced (small $\zeta_j$) thereby decreasing the mode volume.
In this way, SP coupling to atoms increases with increasing SP frequency:
hence the Kerr coefficient and phase shift increase accordingly.

By adjusting the media parameters,
we can achieve giant Kerr nonlinear coefficient and phase shift of the order~$\pi$ at the required frequency, in this case the Rb gas transition frequency corresponding to $\omega=0.144\omega_{\rm e}$.
These results demonstrate clearly that it is possible to achieve the requirements of
low losses, subwavelength confinement, large Kerr nonlinear coefficient and cross phase shift of order~$\pi$.
For coherent state inputs, the lower bound on mean photon number is the desired phase shift,
e.g.\ $\pi$ photons for a $\pi$ phase shift, but this bound could be beat for nonclassical light~\cite{SM92a,SM92b}.

We have also analyzed the solid-state Pr:YSiO system with similar results.
The weak dipole moment of this solid system is compensated for by the large atomic density in typical solids.
The LLSP and their high degree of confinement together with DEIT generate large phase shifts for this system.
However, the Pr:YSiO resonant wavelength of 606nm is a little beyond the reach of current NIMM technology.

\section{Conclusions}
\label{sec:conclusions}
In conclusion, combining SP confinement at a NIMM interface with the DEIT mechanism
yields the trifecta for large cross-phase modulation.
These three sought-after properties are
low loss,
high field confinement, and
large Kerr coefficients.

The goal is to reach mutual phase shifts of~$\pi$ at the single-photon level,
which would have profound implications for quantum information technology.
In the meantime, creating mutual phase shifts of order~$\pi$ for the case of each pulse having
hundreds of photons would be exciting for nanophotonic switches.

With state-of-the-art nanofabrication technology, the nonlinear atomic medium of thickness $z_0$ can be implemented by implanting impurity atoms on the dielectric part of the interface with a NIMM. Although EIT is well studied and experimentally demonstrated in many
labs around the world, and despite the tremendous progress in nanofabrication and NIMM technology, combining EIT and DEIT with NIMMs is the most challenging part in this scheme.
 However, NIMM technology has reached optical frequencies so $^{87}$Rb at the interface with
a NIMM should be feasible soon, and a solid-state implementation with Pr:YSiO viable
as NIMM technology reaches shorter wavelengths.

\acknowledgments
We gratefully acknowledge financial support from {\em i}CORE, NSERC, KACST, and RFBR grant \#08-07-00449.
BCS is a CIFAR Associate

\end{document}